\documentstyle[12pt]{article}

\author{Hao Chen\thanks{Research supported by NNSF}\\
Department of Mathematics\\
Zhongshan University\\
Guangzhou,Guangdong 510275\\
People's Republic of China\\
and\\
Department of Computer Science\\
National University of Singapore\\
Singapore 117543\\
Republic of Singapore}
\title{Some Good Quantum Error-Correcting Codes from Algebraic-Geometric Codes}
\date{April,2000}

\begin{document}

\maketitle
\begin{abstract}
It is shown that the quantum error-correction can be acheived by the using of classical binary codes or additive codes over $F_4$ (see [1],[2],[3]). In this paper with the help of some algebraic techniques the theory of algebraic-geometric codes is used to construct asymptotically good family of quantum error-correcting codes and other classes of good quantum error-correcting codes. Our results are compared with the Tables in [4] and known best quantum codes in [4],[5],[6].

\end{abstract}

\section*{1 Introduction and Preliminaries}

Since the poineering works in [1],[2],[3] now quantum error-correcting codes is rapidly developed and a thorough discussion of the principles of quantum coding theory was offered in [4], and many examples and tables about various bounds were given there. Also many kinds of interesting good quantum codes were constructed by the using of classical binary codes, see,e.g., [5] [6] and [7]. It is natural to consider to use the theory of Algebraic-geometric codes to construct good quantum codes. In this paper, the family of asymptotically good quantum codes is constructed from algebraic-geometric codes over the well-known Garcia-Stichtenoth second tower [8] and some good quantum codes were given by the using of algebraic curves over finite fields of characteristic 2. The general restriction and goodness of the quantum codes from our this construction is also analysised.\\

The paper is organized as follows. We recall basic results of Caderbank-Shor-Steane construction (CSS codes), one theorem in [7] and the basic construction of algebraic-geometric codes (e.g.,see [10]) below. The family of asymptotically good quantum codes is constructed in section 2. In section 3 we construct some good quantum codes from algebraic curves. The conclusions are presented in the last section.\\

We recall the following result in [1], [2] or [4].\\

{\bf Theorem 1.1} (Calderbank-Shor-Steane) {\em Let $C_1$ and $C_2$ be two binary codes with parameters $[n,k_1,d_1]$ and $[n,k_2,d_2]$ respectively. Suppose that $C_1^{\perp} \subset C_2$. Then a quantum $[[n,k_1+k_2-n,min\{d_1,d_2\}]]$ code can be constructed.}\\

Since we use codes over $F_{2^t}$ sometimes instead of binary codes. We need to recall binary expansion $B(C)$ of a linear $F_{2^t}$ code $C$ with respect to a base $B$ of $F_{2^t}$ over $F_2$ as in [7]. It is easy to define the expansion $B: F_{2^t} \rightarrow F_2^t$ with respect to the base $B$. The binary expansion $B(C)$ of the code $C$ is defined componentwise. It is well-known that there exists a self-dual base for the fields of characteristic 2. We need to use the following result of [7].\\

{\bf Theorem 1.2} ([7]) {\em If $B$ is a self-dual base of $F_{2^t}$ over $F_2$. Then the dual $B(C)^{\perp}$ of $B(C)$ is the binary expansion $B(C^{\perp})$ of the dual $C^{\perp}$ of $C$.}\\

Let $X$ be a genus $g$ smooth, projective, absolutely irreducible curve defined over $F_q$, $P$ be a set of $n$ $F_q$-rational points of $X$ and $G$ be a $F_q$-rational divisor of $X$, such that, $supp(G) \cap P =\emptyset$ and $2g-2<deg(G)<n$. Then the functional $C_L(G,P)$ and residue $C_{\Omega}(G,P)$ algebraic-geometric codes can be defined. It is well-known that the dual of $C_L(G,P)$ is $C_{\Omega}(G,P)$ (e.g.,see [10]Cha.II, section 2).There parameters are as follows.\\

{\bf Theorem 1.3}(e.g.,see [10] II.2) {\em The functional code $C_L(G,P)$ is a $[n,deg(G)-g+1,n-deg(G)]$ linear code over $F_q$ and the residue code $C_{\Omega}(G,P)$ is a $[n,n-deg(G)+g-1,deg(G)-2g+2]$ linear code over $F_q$.}\\

\section*{2 Asymptotically good quantum codes}

Now we recall the results in [8](see section3 of [8]) about a tower of function fields $T_1 \subset T_2 \subset T_3....\subset T_i \subset T_{i+1}...$ over $F_{q^2}$ attaining the Drinreld-Vl\u{a}du\c{t} bound,i.e., $\lim sup (N(i)/g(i))=q-1$, where $N(i)$ and $g(i)$ are the number of $F_{q^2}$-rational points and genus of $T_i$ respectively. These function fields are given by $T_1=F_{q^2}(x_1),...,T_{i+1}=T_i(x_{i+1})$, with $x_{i+1}^q+x_{i+1}=x_i^q/(x_i^{q-1}+1)$. Let ${X_i}$ be the family of algebraic curves over $F_{q^2}$ corresponding to the function fields ${T_i}$. Let $\Omega$ be the set of $q$ roots of $y^q+y=0$ in $F_{q^2}$. We note that for each value of $x_1$ in $F_{q^2} \setminus \Omega$, $x_1^q/(x_1^{q-1}+1)=x_1^{q+1}/(x_1^q+x_1)$ is in the set $F_q \setminus {0}$, thus we have $q$ distinct solutions of the equation $Tr_{F_{q^2}/F_q}(x_2)=x_2^q+x_2=x_1^q/(x_1^{q-1}+1)$ in $F_{q^2}$ and each such solution $x_2$ is in the set $F_{q^2} \setminus \Omega$. Hence we can continue this argument to get at least $(q^2-q)q^{i-1}=(q-1)q^i$ $F_{q^2}$-rational points of the curve ${X_i}$ (see [8] Lemma 3.9 in p.265). Moreover it is known that the genus $g(2i)=(q^{i}-1)^2$ and $g(2i+1)=(q^{i+1}-1)(q^{i-1}-1)$ (see [8],p.265).\\

From now on we take $q=2^t$. We take two points $P^{(h)},Q^{(h)}$ from these $(q-1)q^h$ $F_{q^2}$-rational points and let $D_h$ be the set of other $(q-1)q^h-2$ $F_{q^2}$-rational points of $X_h$. Consider algebraic-geometric functional code $T_1=C_L(mq^h P^{(h)},D_h)$ with $2 <m <q-1$. It is a $[\#D_h, mq^h-g(h)+1,\#D_h-mq^h]$ linear code over $F_{q^2}$. Its dual is the algebraic-geometric residue code $T_1^{\perp}=C_{\Omega}(mq^hP^{(h)},D_h)$. We know the residue code $T_2=C_{\Omega}((mq^hP^{(h)}-q^hQ^{(h)}),D_h)$ contains $T_1^{\perp}$ as a subcode. It is a  $[\#D_h,\#D_h-(m-1)q^h+g(h)-1,(m-1)q^h-2g(h)+2]$ code (see Theorem 1.3).\\

{\bf Theorem 2.1} {\em There exists a family of quantum $[[n_h,k_h,d_h]]$ codes \{$C_h$\}, such that, $R=\lim (k_h/n_h)$ and $\delta=\lim (d_h/n_h)$ are positive and $R+\delta \geq \frac{1}{12}$.}\\

{\bf Proof.} We take the binary code $C_1=B(T_1)$ and $C_2=B(T_2)$, where $B$ is any given self-dual base of $F_{2^t}$ over $F_2$. From Theorem 1.2 $C_1^{\perp}=B(T_1^{\perp})$ is a subcode of $C_2$. Thus the condition of Theorem 1.1 is satisfied. We have a quantum $[[n_h,k_h,d_h]]$ code with\\

$$
\begin{array}{cccccc}
n_h=2t((2^t-1)2^{th}-2);\\
k_h=2t2^{th};\\
d_h=min\{(2^t-1-m)2^{th}-2,(m-3)2^{th}\}
\end{array}
(2.1)
$$

Hence $R=lim (k_h/n_h)= \frac{1}{2^t-1}$ and $\delta= \frac{min\{2^t-1-m,m-3\}}{2^t-1}$. From the restriction on $m$ we have to take $t \geq 3$, thus, $R= \frac{1}{2^t-1}$ and $\delta \geq \frac{2^{t-1}-3}{2t(2^t-1)}$, when $m=2^{t-1}$.We know that $R$ and $\delta$ are positive and $R+\delta  \geq  \frac{1}{4t}$. Thus the conclusion is proved when we take $t=3$.\\

From Theorem 2.1 the family of asymptotically good quantum codes is constructed.\\

\section*{3 Quantum codes from curves over $F_{2^t}$}

We consider the quantum codes from Theorem 1.1 by the using of algebraic curves over $F_{2^t}$.\\

{\bf Theorem 3.1} {\em Suppose $X$ is an algebraic curve defined over $F_{2^t}$ with $N$ rational ponits and genus $g$. Let $m,m'$ be positive integers satisfing $2g-2<m<N$ and $0 \leq m'<m-2g+2$. Then there exist a quantum $[[t(N-2),tm',min\{N-2-m,m-m'-2g+2\}]]$ code.}\\

{\bf Proof.} We take 2 rational points $P,Q$ of $X$ and let $D$ be the set of other $N-2$ rational points. Consider the code$T_1=C_L(mP,D)$ and $T_2=C_{\Omega}((mP-m'Q),D)$. From the well-known fact in algebraic-geometric code theory (see [10]), we have $T_1^{\perp}=C_{\Omega}(mP,D)$ is a subcode of $T_2$. Let $B$ be any self-dual base of $F_{2^t}$ over $F_2$, $C_1=B(T_1)$ and $C_2=B(T_2)$. From Theorem 1.2 we have $C_1^{\perp}=B(T_1^{\perp}) \subset B(T_2)=C_2$. Thus the condition of Calderbank-Shor-Steane code (Theorem 1.1) is satified. It is clear $C_1$ is a  $[t(N-2),t(m-g+1),N-2-m]$ binary code and $C_2$ is a $[t(N-2), t(N-2-(m-m')+g-1),m-m'-2g+2]$ binary code from the corresponding parameters of functional and residue algebraic-geometric codes(see Theorem 1.3). Thus from Theorem 1.1 we get our conclusion.\\

This result offers a lot of quantum codes with flexibility of choices of parameters as in classical algebraic-geometric code theory (see [10] and comparing to quantum codes in [4],[5],[6]). For the existence of algebraic curves with specifc $(N,g)$ we refer to G. van der Geer and M. van der Vlugt's list [9]. For example, if we use an algebraic curve over $F_4$ with $(N,g)=(9,1)$ (ie. maximal elliptic curve over $F_4$)we get $[[14,0,3]], [[14,2,3]], [[14,4,2]]$ $[[14,6,2]], [[14,8,1]]$ quantum codes.When we take curve over $F_4$  with $(N,g)=(17,5)$ in [9], we get $[[30, 0, 3]], [[30, 2, 3]], [[30, 4,2]], [[30, 6,2]],[[30,8,1]]$ quantum codes.\\

From Theorem 3.1, the projective curves over $F_{2^t}$ with $(N,g)=(2^t+1,0)$ can be used to construct quantum codes.\\

{\bf Corollary 3.2} {\em Let $t,m,m'$ be positive integers with $m<m'<2^t-1$. We have a $[[t(2^t-1),tm',min\{2^t-1-m,m-m'+2\}]]$ quantum codes.}\\

{\bf Proof.} This is direct from Theorem 3.1.\\

It is well-known that maiximal elliptic curve,ie., curves with $(N,g)=(4^t+2^{t+1}+1,1)$, exist for any field $F_{2^{2t}}$ (e.g., see [11]). We have the following result.\\

{\bf Corollary 3.3} {\em Let $t,m,m'$ be positive integers and suppose $m'<m<4^t+2^{t+1}-1$ Then there exists a $[[2t(4^t+2^{t+1}-1),2tm',min\{4^t+2^{t+1}-1-m,m-m'\}]]$ quantum codes.}\\

{\bf Proof.} This is direct from Theorem 3.1.\\

Hermitian curves(see [10], VII.4) over $F_{2^{2t}}$ with $(N,g)=(8^t+1,(2^t-1)2^{t-1})$ can be used to construct quantum codes as in the following statement.\\

{\bf Corollary 3.4} {\em Let $t,m,m'$ be positive integers with $2^t(2^t-1)-2<m<8^t-1$ and $m'<m-2^t(2^t-1)+2$. We have a $[[2t(8^t-1),2tm',min\{8^t-1-m,m-m'-2^t(2^t-1)+2\}]]$ quantum code.}\\

{\bf Proof.} This is direct from Theorem 3.1.\\

\section*{4 Conclusions}

After computing the short quantum codes constructed from Theorem 3.1 by the using of curves in the list in [9], it seems that these quantum codes cannot reach the highest possible minimum distance in the TableIII of [4]. However as the  example given in section 2 we can get a family of asymptotically good quantum codes from algebraic-geometric codes over asymptotically good tower of curves over $F_{64}$. This phonomenon is similar to the case of classical codes from algebraic-geometric construction. However we think that it is possible to get good lower bound of minimum distance of the binary expansion of algebraic-geometric codes, and thus it seems possible better short quantum codes may be constructed.\\

\begin{center}
REFERENCES
\end{center}

1. A.R. Calderbank and P.W.Shor, "Good quantum error-correcting codes exist," Phys.Rev. A, vol.54, pp1098-1105, Aug.,1996\\

2.A.M.Steane, "Multiple particle interference and quantum error correction, Proc. Roy. Soc. London, A, vol.452, pp2551-2577, Nov.,1996\\

3. P.W. Shor, "Scheme for reducing decoherence in quantum computer memory", Phys. Rev., A, vol.52, ppR2493-R2496, Oct.,1995\\

4. A.R.Calderbank, E.M. Rains, P.W.Shor and N.J.A. Sloane, "Quantum error correction via  codes over GF(4),", IEEE Trans. Inform. Theory, vol.44, pp1369-1387, 1998\\

5. A.M. Steane, "Enlargement of Calderbank-Shor-Steane quantum codes", IEEE Trans. Inform. Theory, vol.45, pp2492-2495, Nov.,1999\\

6. G. Cohen, S. Encheva and S.Litsyn, On binary construction of quantum codes", IEEE Trans. Inform. Theory, Vol.45, pp.2495-2498, Nov.1999\\

7. M.Grassl,  W.Geiselmann and T.Beth, "Quantum Reed-Solomon codes", in Proceedings of AAECC-13, M.Fossorier, H.Imai, S.Lin and A.Poli (Eds), LNCS1719, Springer-Verlag, 1999, pp231-244\\

8. A.Garcia and H. Stichtenoth, On the asymptotic behaviour of some towers of function fields over finite fields, J. Number Theory, vol.61, pp248-273, 1996\\

9. G. van der Geer and M.van der Vlugt, Tables of curves with many points, updated Feb.17, 2000 \\

10.H.Stichtenoth, Algebraic function fields and codes, Universitext, Springer-Verlag, 1993\\

11.M.A.Tsfasman and S.G.Vladut, Algebraic-geometric codes, Kluwer Academic Publishers, Dordrecht-Boston-London, 1991\\

{\bf Hao Chen} received the Ph.D from Fudan University, Shanghai, China in 1991. From 1991-1993, he was a postdoctoral researcher in Computing center, Academia Sinica (Beijing). Now he is with Department of Mathematics, Zhongshan University, Guangzhou, China and Department of Computer Science, National University of Singapore, Republic of Singapore. His research fields include algebraic-geometric codes, codes over Galois rings, quantum error-correcting codes and singularities of algebraic varieties.

\end{document}